\documentclass[10pt,conference]{IEEEtran} 

\usepackage[T1]{fontenc}
\usepackage[latin1]{inputenc}
\usepackage[tight,footnotesize]{subfigure}
\usepackage{picins}

\usepackage{ifpdf}
\ifpdf
    \usepackage[pdftex]{graphicx}
 \else
  \usepackage[dvips]{graphicx}
\fi

\newcommand{\hamcast}{\mbox{\sf H\boldmath$\forall$Mcast }}

\IEEEoverridecommandlockouts

\begin{document}

\title{Why We Shouldn't Forget Multicast in Name-oriented Publish/Subscribe}


\author{
\IEEEauthorblockN{Thomas C. Schmidt}
\IEEEauthorblockA{HAW Hamburg,
Dept. Informatik\\
Berliner Tor 7\\
D--20099 Hamburg, Germany\\
Email: t.schmidt@ieee.org}
\and
\IEEEauthorblockN{Matthias W{\"a}hlisch}
\IEEEauthorblockA{FU Berlin,
Inst. f{\"u}r Informatik\\
Takustra{\ss}e 9\\
D--14195 Berlin, Germany\\
Email: waehlisch@ieee.org}
}

\IEEEaftertitletext{\vspace{-0.55cm}}
\maketitle

\begin{abstract}
Name-oriented networks introduce the vision of an information-centric, secure, globally available publish-subscribe infrastructure. Current approaches concentrate on unicast-based pull mechanisms and thereby fall short in automatically updating content at receivers. In this paper, we argue that an inclusion of multicast will grant additional benefits to the network layer, namely efficient distribution of real-time data, a many-to-many communication model, and simplified rendezvous processes. These aspects are comprehensively reflected by a group-oriented naming concept that integrates the various available group schemes and introduces new use cases. A first draft of this name-oriented multicast access has been implemented in the \hamcast middleware.
\end{abstract}

\begin{IEEEkeywords}
Content Centric Networks, Multicast, Naming, Addressing, Security, Routing
\end{IEEEkeywords}

\section{Introduction}

The search for future directions in Internet development has drawn significant attention to name-based communication concepts that are built upon the publish/subscribe paradigm. Inspired by the Web use case and widely deployed content distribution networks, the proposed {\em Content} or {\em Information Centric Networks} (CCN/ICN) abandon the current Internet model of connecting end nodes. Instead, consumers shall retrieve content by name directly from a network that provides storage, caching, content-based rendezvous, and searching at times.

Several proposals have been presented in recent years, among them TRIAD \cite{gc-acrsi-01}, DONA \cite{kccek-dona-07}, NDN \cite{zebjt-ndnp-10,jstp-nnc-09}, PSIRP/LIPSIN \cite{jzran-llsps-09}, and NetInf \cite{a-snad-10}. The schemes differ in naming/addressing, routing/rendezvous, security/authentication, forwarding/caching, and minor design choices, but jointly consider Multicast at most as a side property of data paths. Only very recently, COPSS \cite{cajfr-cecop-11} has brought  up the discussion of relevance for a group distribution model in CCN. COPSS targets at seamlessly extending NDN by a specific multicast service based on PIM-type rendezvous points, thereby omitting a broader discussion about naming, addressing, routing and security. The present paper attempts to fill this gap. 

Multicast is the traditional approach to publish/subscribe on the Internet layer \cite{dc-mrdie-90}. In contrast to CCN networks, which also facilitate the consumption of a single (cached) data copy by multiple receivers, multicast enforces a highly efficient, scalable data forwarding which is even used in MultiCache \cite{kxp-moain-11} to improve the CCN efficiency.
In addition, the multicast model features the following relevant aspects. 

\begin{itemize}
\item Data is pushed to receivers, supporting multiple transport streams in parallel and eliminating the need to ask for content changes.
\item Data distribution supports immediate in-network forwarding and is suitable for efficient, scalable real-time streaming in particular. This mainly covers the use case of real-time date dissemination without storage or caching requirements.
\item The multicast model contributes many-to-many communication, which is valuable whenever information is created at distributed origins. Multi-source communication using a single name faces strong conceptual difficulties in unicast-based CCNs.
\item Multicast group communication enables rendezvous processes, as publishers and subscribers remain decoupled and unknown to each other.
\end{itemize} 

In this discussion paper, we collect arguments for the case of multicast communication in name-oriented pub/sub networks. Led by the observation that a majority of today's applications rely on group communication,\footnote{Group communication today is - similar to content distribution - almost always implemented on the application layer.} we start with a discussion of various multicast aspects that are missing in the current CCN models. In the core, we continue with a thorough conceptual examination of naming, addressing, grouping and security for multicast communication in information-centric networking. We present an integrative proposal for a common multicast access scheme that donates particular attention to the programming side, as well as an overview of the various routing options.

Our naming concepts that include hierarchies and aggregation as well as named instantiations comply to the principles of ``push enabled dissemination'', ``decoupling of publishers and subscribers'', and ``support hierarchies and context in naming content'' as postulated in \cite{cajfr-cecop-11}. In addition, we are able to express further group concepts such as selective broadcast and selective data origins, and clearly separate content from node names in our syntax. The latter is of particular importance for building conceptually clear, type-safe programming models on top of our networking concept.

The remainder of this paper is structured as follows. In Section \ref{sec:problemspace} we discuss the problem space and identify the unique contributions the multicast model can add to the CCN paradigm. Section \ref{sec:naming} is dedicated to naming and its implications for group forming, routing and security. A summary on available routing options is presented in Section \ref{sec:routing}. Finally, Section \ref{sec:c+o} concludes on the achievements, provides reference to our implementation and previews on the upcoming steps in this ongoing work.

\section{Why do We Need Multicast in\\ Name-oriented Networks?} \label{sec:problemspace}

CCN/ICN-style networks introduce the vision of a secure, efficient,
globally available publish/subscribe system. In the current Internet, the
publish/subscribe paradigm has been implemented by multicast, even though
it is not globally deployed, yet, and only present at the intra-domain level.
This paradigmatic coincidence raises two questions: What are the differences between multicast and content centric networks in detail? Why should we should care about multicast on the CCN layer?


\subsection{Recap the Multicast Model}

The multicast communication model implements the following functions:
\begin{enumerate}
  \item Explicit data subscription by receivers.
  \item Delocalized content storage.
  \item Content replication per group.
  \item Open group as well as closed group concept.
\end{enumerate}

A multicast listener must join a multicast group to receive the
corresponding data. Based on the subscription, distribution paths will be
established dynamically. Thus, the network infrastructure delivers
multicast data only to those nodes that requested the data explicitly. This
is the nature of publish/subscribe systems and in contrast to current
unicast communication, in which a sender may transmit data to parties on the
network layer that never agreed on the communication.

In principle, the multicast group model splits up in one-to-many (Source
Specific Multicast) and many-to-many multicast (Any Source Multicast). The
latter is open per se, while the first allows for closed groups. Securing
multicast group management has been discussed since a while
\cite{cbb-tmdoa-04}, even for the mobile regime, where self-certifying
group identifiers have been proposed to prevent a distribution of content
by illegitimate sources \cite{swch-amcpf-08}.

Any source multicast inherently provides functions to address content by an
identifier that is fully decoupled from the actual storing host. On the
application side, there is no direct binding between content ID and host
address of the source. Content can be provided by one or multiple arbitrary
sources, while receivers need not resolve the location of content.

From both perspectives, receiver-driven content access and content naming
instead of addressing, multicast and content-centric networking follow the
same paradigm. However, there is an important difference in how content is
duplicated within the network.

Content replication in multicast is initiated by the receivers. Multicast
on the data link, network, and transport layer is restricted to unbuffered
push. Intermediate nodes do not store content in advance or cache the
content for subsequent receivers. In contrast, CCN/ICN initiates
data replication by the network itself. Early previous work in the context of web caching considers multicast conjointly with consumer-oriented content placement.  In the extreme case, a web server would continuously transmit  
content to a multicast group that will be joined by WWW clients. However,
this is only efficient for continuous content and may be
generalized by push-caching schemes~\cite{rrb-icm-98}.




\subsection{Ease Content Replication}

Content replication describes the process to disseminate the same data to
different places. The content may be delivered directly to the consumers or
stored at pre-located content repositories. Replication provides a
consistent view on the distributed content. In this sense it differs
from caching, which applies local replacement strategies and thus may lead
to inconsistencies. Network layer multicast implements an efficient way to
replicate content towards interested peers.

\begin{figure}
  \subfigure[Repeated access:  On-demand replication close to the
  consumer, data caching for further access.]{\includegraphics[height=2.6cm]{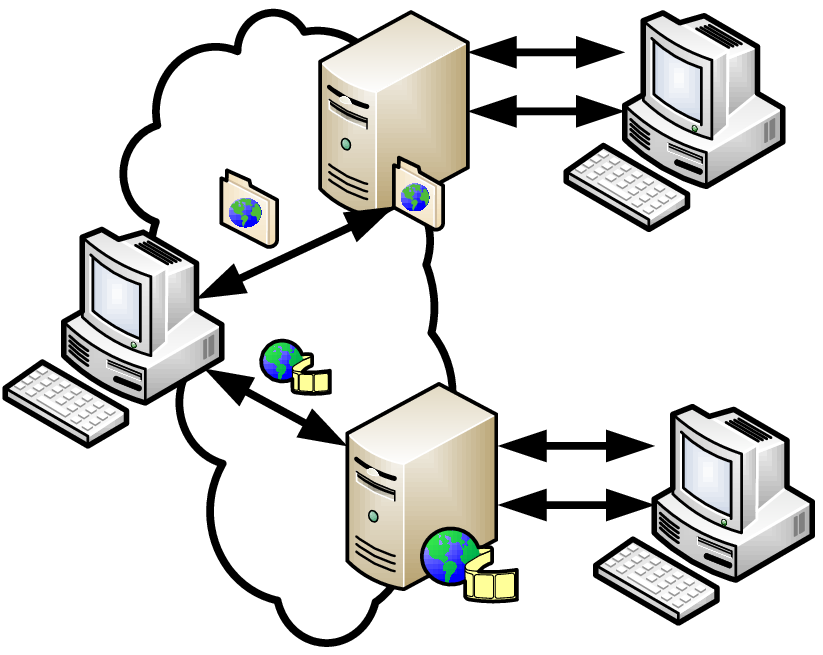}}\quad
  \subfigure[One-time access: Storage in advance close~to~the~consumer, 
  but data can be removed after access.]{\includegraphics[width=0.14\textwidth]{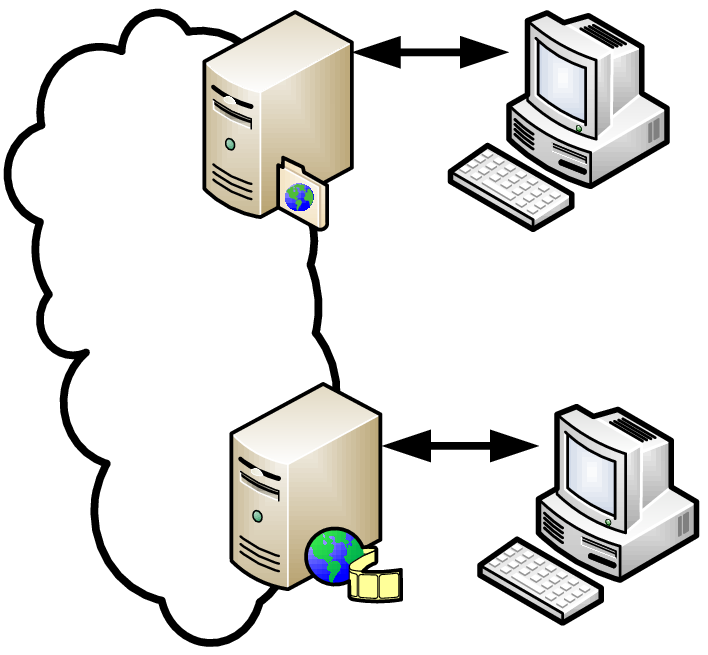}}\quad
  \subfigure[Synchronous multi-party access: Storing of content in
  the vicinity of a group of consumers.]{\includegraphics[width=0.14\textwidth]{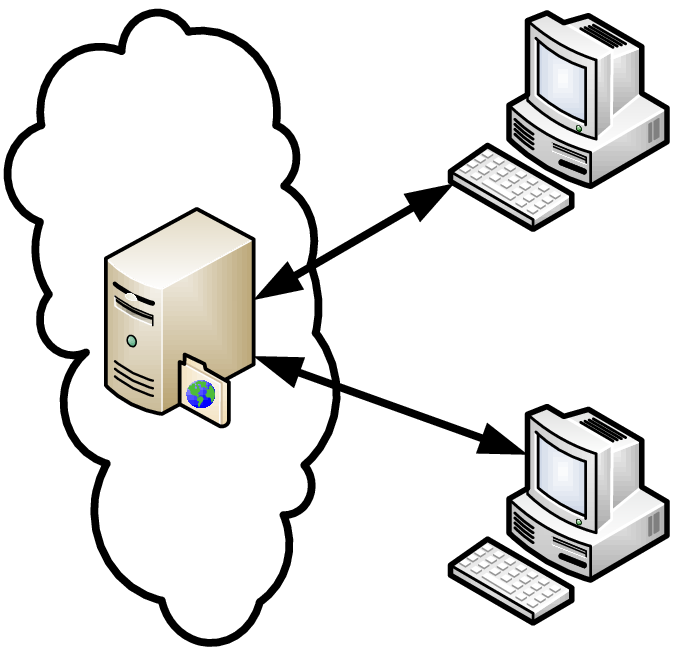}}
  \caption{Different content replication scenarios}\label{fig:replication}
\end{figure}

Network-based content replication is motivated by two reasons: the
reduction of network costs (i.e., eliminate redundant data transmission)
and the improvement of the end user experience (i.e., decrease delay).
There are three generic application scenarios, in which this is helpful (see 
Figure~\ref{fig:replication}). In the first case, a single node accesses
the content multiple times. This may occur due to limited local buffers at
lightweight mobile nodes, for example. To achieve both goals of content
replication, the data should be stored as close as possible to the receiver
and as long as possible to avoid unnecessary retransmissions.

The second case describes a host that downloads content one-time. The
delivery should be as fast as possible. In contrast to the previous
scenario, the content can be deleted immediately after the access to
release memory resources, but must be available in
advance to diminish network delay.

If multiple nodes request the same content in parallel (i.e., a case for multicast), the content needs replication at several branch points. From a network efficiency perspective, such replication should not be simply performed close to content consumer, but focus on a group perspective. From the alternative storage point of view, content should be placed in the vicinity of the receivers.

In all three scenarios, an optimal replication strategy depends on the use
case. To prevent repeated transmissions of the same content, a distributed
storage is important for long-living data. The subset of participating
peers determines the distance between content repository and the consumers.
An advanced knowledge about the type of content (multicast vs.
unicast), thus, may help to improve the underlying distribution
mechanism. As already noted in \cite{rrb-icm-98}, a subscription-oriented
model eases the identification of content that is of interest for receivers
and so improves pre-fetching of the data.  Named-data networking may
benefit from an explicit consideration of multicast, as this enables to
adjust replication strategies,and thus reduces complexity and saves costs.

%








\subsection{Synchronous Content Access}

Multicast describes a distribution mechanism is closely bound to 
specific content types. Remaining indifferent to multicast and
unicast content reduces the ability to consider the underlying semantic 
at the application site. Multicast content is \emph{synchronously} accessed
by multiple receivers. Any receiver who is interested in the content gets
the same (sub-)piece of data at the same time, even if the content
descriptor refers to a set of fragments. A video stream, which naturally
consists of multiple sequences, may serve as one example. The pub/sub paradigm of CCN/ICN grants sufficient flexibility to allow for the identification of any
single piece of continuous content, but requires polling of each data packet \cite{jsbps-voccn-09}. 

Any abstraction from the packet level in CCN/ICN creates a confusing compexity. For exanmpe, a dedicated scene may be named by
\texttt{time.movie.cnn.com}. However, accessing this content leaves the
outcome ambiguous: Either the video starts playing or the consumer jumps
into a subsequent scene. As long as there is no differentiation between
multicast and non-multicast content, only one option can be implemented, and
there is no meaningful default behavior.

\subsection{Application Programming}

The programming of multicast applications follows a different perspective as 
compared to unicast. In multicast, a programmer opens a socket, subscribes
to content, and waits for an arrival of data for the time of interest. An application program thus decouples the time of data reception from the instance of its readiness, but attempts to synchronize to transmission. In the 
case that content is not available during the subscription period, the application programmer does not experience an error, neither seeing the distribution system at fail, nor the application layer. In contrast, when a chunk or file is
accessed in unicast communication, the connection end point experiences an
error whenever the content is unavailable. The networking layer triggers an
unreachable message for a connection that cannot be established, or the
application layer  typically responds with a failure messages (e.g.,
HTTP~404).

The availability of multicast content cannot easily be verified. The
content consumer is not aware of any multicast source prior to data arrival,
and thus cannot check whether a potential content source is available. The
absence of data does not indicate the absence of a content source, as there
are several multicast applications that announce data only sporadically
(e.g., NTP). The current networking layer is also not aware of multicast
sources on a global scale. Broadcast-like routing schemes such as DVMRP are not
suitable for inter-domain routing. On the other hand, protocols based on rendezvous points such as PIM-SM do not inform participating routers about the
arrival of a new source.

Content-centric networks may or may not operate similar to multicast and omit or provide an explicit error signaling on the distribution layer. Only the corresponding function can then be reflected at the content programming interface.
Consequently, an application programmer needs to handle one (the alternate) case on his own (e.g., by timeouts). However, this burdens complexity to the implementation site. The support of both mechanisms, pull-oriented unicast and push-awaiting multicast, is indeed desirable.

\section{Naming and Addressing in Multicast}\label{sec:naming}

Traditionally, naming and addressing in multicast-type group communication follow technology. Group addresses in IP that serve the ASM model have never successfully entered DNS, while no naming standard supports source-specific $(S,G)$-channels. Similarly, overlay and application-layer schemes have been built on diverse hashing or naming conventions that remain bound to specific use of routing or application processing. At present, {\em programmers} need to decide on deployment technologies by selecting types for names or addresses.

Multicast today does not involve a clear concept of addressing. On the Internet layer, `multicast addresses' are de-localized parameters for routing states, unless unicast route information is included for sources \cite{hc-imces-99} or rendezvous \cite{RFC-3956}. A similar picture is visible for application layer multicast, where rendezvous \cite{rkcd-sdlse-01}, instantiation \cite{rhks-amucn-01} or reflector services are the only use of location semantics in multicast group names. Following such observations, John Day coined ``Multicast addresses is a set of
distributed application names''~\cite[p.  329]{d-pna-08}. Up until now, traditional networking has failed to deliver a uniformly applicable naming and addressing concept for groups, as they do not represent routing endpoints. The newly aligned perspective of a general publish-subscribe paradigm in networking may encourage to undertake this approach  anew.

\subsection{Design Aspects for Multicast Names}

Starting from the plain objective of naming groups, one could be tempted to opt for simple string identifiers. However, the design of a naming scheme for multicast groups bears a number of specific aspects that go beyond flat strings or unicast-type content subscription. We discuss core aspects in the following.

\subsubsection{Coverage of Group Semantics}

A uniform naming scheme for multicast should cover the large variety of prevalent group semantics. In particular, an application programmer should not be forced into selecting an interface or data type that predefines the kind of group management or distribution technique. Multicast names, instead, need to provide the expressiveness of all multicast flavours within a single \mbox{(meta-)} data type. Application code including user interfaces may thus remain transparent with respect to the distribution system, while its corresponding intelligence can be established at the end system or network level.
   
\subsubsection{Namespace Support}

The heterogeneity of multicast semantic, but also diverse aspects of routing and transport request for a variable interpretation and processing of group names.  For example, the routing system may need to contact a mapping service in some instances, in others  multicast data may be accessible via distributed content replication servers (e.g., a CDN) that are expressed as a set of instantiations. A name may further hint to an address mapping in a default technology, or indicate a cryptographic algorithm. 

In all these cases, the network, the end system, or the application are required to identify and process the encodings and operate accordingly. To allow for an unambiguous interpretation and a type-safe implementation, indicators of namespaces are of versatile use.

\subsubsection{Instantiation}

The model of source-specific multicast  restricts subscriptions to single-source $(S,G)$ states for the sake of a simplified routing. It resolves the rendezvous problem of ASM in the special case of a single publisher. SSM-type service instantiation should be supported by a uniform naming scheme, but in addition may be extended to cover more general cases. 

A group communication system can be instantiated not only by a single source, but by a set of originators that either act (a) as content replicators following an anycast semantic, or (b) as a source-group representing an explicitly named or implicit many-source semantic, or (c) as a remote distribution system (e.g., an overlay) that discloses multicast content on some membership contact. 

A corresponding syntax, e.g., of the form {\tt <group>@ <instantiation>}, must comply with this rich set of semantics. The clause {\tt <instantiation>} can yield this expressiveness (i) as an indirection by pointing to some mapping service or rendezvous node, or (ii) by referring to a bootstrap point for contacting a remote overlay, or (iii) by explicitly enumerating a set (e.g., {\tt $\{inst_1,...,inst_n\}$}, or (iv) by implicitly naming a set in the form of a statistical filter (e.g., a Bloom filter \cite{b-stthc-70}). 

While an instantiation is part of the logical identifier of a multicast group (e.g.,  {\tt news@cnn.com} and {\tt news@bbc.co.uk} name two different groups), the proposed syntax clearly distinguishes namespaces and semantics. The {\tt <group>} clause names content without reference to a network endpoint, whereas  {\tt <instantiation>} refers to a (group of) publishing node(s).
Much like in the IP/SSM model, naming (sets of) instantiations can guide the routing layer, an end system, or an application to steer pub/sub contact messages. We note that false positives induced by our use of Bloom filters can lead to unwanted contact paths, which some distribution technologies may mitigate by negotiating pub/sub  with the instantiation nodes in a two way handshake.

\subsubsection{Hierarchy and Aggregation}

Hierarchical naming introduces aggregation, which bears an inherent concept of grouping. The corresponding expressiveness of names gives rise to a number of group applications. 

A selective broadcast may for instance become accessible by `wildcarding': Suppose there are news channels {\tt politics@cnn.com}, {\tt economics@cnn.com}, etc. Simultaneous publishing to all (possibly unknown) channels may be enabled via {\tt *@cnn.com}. 

For a subscriber-centric example, consider a layered video stream {\tt blockbuster} available at different qualities  {\tt Q$_i$}, each of which consist of the {\tt base} layer plus the sum of {\tt EL}$_j, j \leq i$ enhancement layers. Each individual layer may then be accessible by a name {\tt EL$_j$.Q$_i$.blockbuster, $j \leq i$}, while a specific quality aggregates the corresponding layers to {\tt Q$_i$.blockbuster}, and the full-size movie may be just called {\tt blockbuster}.

It may be useful to aggregate instances of publications, as well. Multiple news channels, for example, may be available from {\tt news@cnn.com, news@bbc.co.uk,...}. A subscriber may wish to select multiple channels jointly expressed in a set of sources, or request for all news channels using the group aggregator {\tt news}. While the latter step resembles the transition from SSM to ASM, it is worth noting that a growing number of sources (aggregated in a set or in a Bloom filter) steadily reduces the specificity of the instantiation and thereby implies a smooth transition from the SSM to the ASM multicast model.

Jointly operating on the identifiers for groups and instantiations, this name-aggregation concept provides rich expressiveness with heterogeneous deployment options. We should emphasize that it transparently abstracts from a particular service deployment as a multi-source or single-source multicast communication.

\subsubsection{Canonical Support for Stateless Mapping}

Many different distribution technologies for multicast are deployed today, and heterogeneity may be expected to persist in future content centric networks. Some technologies use group identifiers of a specific syntax such as IP multicast addresses, overlay hash values, or SIP group conference URIs. A creator of a group may want to express the desire of using a dedicated group ID in a deployment (s)he prefers. The multicast naming scheme should provide a corresponding expressiveness that enables a stateless canonical mapping of group names to technological identifiers. 

For example, {\tt mcast-ip://224.1.2.3:5000} indicates the correspondence to {\tt 224.1.2.3} in the default IPv4
   multicast address space at port {\tt 5000}.  This default mapping is bound
   to a technology and may not always be applicable, e.g., in the case
   of address collisions.

\subsection{Security in Multicast Naming}

Multicast content distribution requires measures to ensure the four common components of network security: integrity, confidentiality, provenance, and availability \cite{gkrss-nca-11}. Unlike in unicast, though, group communication imposes the following additional constraints. 

\begin{enumerate}
\item Multicast publishers and subscribers are decoupled and commonly allow only for unidirectional signaling. This conflicts with cryptographic handshaking as frequently used to create confidentiality or authenticity.
\item Multiple sources may contribute to the content of a group, making it difficult to prove original and legitimate publishers (provenance).
\item Streaming is a common multicast application. Rather than plain content hashes, the use of stream ciphers is required to ensure integrity.
\item The network infrastructure  assists multicast distribution (even) more strongly than unicast-based content centric networks. Data replication services at a possibly large scale raise well-known threats to the infrastructure and the end systems. Admission control and infrastructure protection are required to ensure availability.
\end{enumerate}

Multicast names are commonly created by sources (publishers) that somehow have gained admission to inject content streams into the network. It is thus reasonable to use the source identification as a trust anchor \cite{swch-amcpf-08} when generating self-certifying names \cite{mkkw-skmfs-99} of multicast content. 

In detail, the creator (controller) of a group that has generated its cryptographic ID from a public-private key pair $({\cal K}_{pub},{\cal K}_{sec})$, will use ${\cal K}_{pub}$ to configure a group name $G$ equally as a cryptographic identity. Conflicts within the node ID space can be avoided by adding a counter. In signing content using ${\cal K}_{sec}$ and attaching ${\cal K}_{pub}$,  the group controller will provide  cryptographically
strong proof of ownership to any receiver of the packet. Each intermediate router (or receiver) can verify the source-group-content relationship after extracting ${\cal K}_{pub}$ and validating the signature.  Assuming that access permission has been granted at the network edge, multicast replication services can be consequently bound to an autonomous packet authentication without feedback loop.

Multiple multicast senders contributing to the same multicast group require
admission by the group controller. This admission
authority has created the cryptographic group name. Before an additional
multicast source $S$ injects data, it requests a certificate. The group controller authenticates the sender and -- according to an application policy -- issues the
certificate, which includes $S$,  the peer membership of
$G$ and an optional lifetime. The certificate is signed with the private key corresponding to the creation of $G$.
A multicast source that wants to transmit data attaches this certificate
and signs packets with its own private key. An intermediate router verifies whether the
group certificate is valid and the group address $G$ has been generated from the
group public key. Additionally, the router authenticates the source's cryptographic identity 
 according to the certificate and the peer  identifier as described in the single-source case.

Multicast content streams require stream cyphers to be linked with the cryptographic identity, the details of which are beyond the scope of this article, see \cite{cbb-tmdoa-04} for a general overview.

\subsection{A Naming Scheme for Multicast}\label{sec:urischeme}

Following our previous design discussion, we now summarize an URI-based naming scheme (cf., \cite{draft-common-api}). 

\begin{verbatim}
  scheme "://" group "@" instantiation ":"
               port "/" sec-credentials
\end{verbatim}

The \emph{scheme} refers to the specification of the assigned identifier
(e.g., mcast-ip, sip, ...), \emph{group} denotes the group ID, \emph{instantiation}
identifies the entity that generates the instance of the group, \emph{port}
identifies a specific application, and \emph{sec--credentials} add optional
cryptographic identities, see \cite{draft-farrell-decade-ni-00} for a corresponding approach. Valid group IDs will be mcast-ip://224.0.1.1:4000 and
sip://hypnotic-talks@psychic.org, for example.

The proposed syntax of the group name provides a consistent term for ASM as
well as SSM-type groups on the application layer. For example,
\emph{mcast-ip://224.10.20.30@1.2.3.4/groupkey} describes an SSM group
name, where \emph{mcast-ip://224.10.20.30/groupkey} denotes a
source--independent multicast group.

The syntax also allows for flexible namespace support.  The group name is
defined by the multicast source. A multicast source is enabled to indicate
a preferred forwarding scheme using a namespace that corresponds to a
distribution technology.

Along this line, a multicast application developer opens a URI--aware
multicast socket without predefining the distribution technology. In the
case of a receiver subscription, for example:
\begin{verbatim}
  ms=createMSocket() 
  ms.join(URI(
    "mcast-ip://224.10.20.30@1.2.3.4\. 
             /groupkey"
  )).
\end{verbatim}

The socket can be used for another multicast group, as well:
\begin{verbatim}
  ms.join(URI(
    "sip://hypnotic-talks@psychic.org"
  )).
\end{verbatim}

\section{Routing to Groups}\label{sec:routing}

Multicast content is targeted at a group of receivers that is partially unaware of the distribution system. In the following, we sketch the options of transporting data from (a group of) sources to the receivers.

\subsection{Flooding Groups of Receivers}

Content may be distributed as broadcast to a group of consumers or replicators (i.e., content caches) following a group building process. Such operations are known from DVMRP on the IP-level or application-layer multicast on CAN. All members of the distribution system thereby share the same data. 
 
\subsection{Reverse Path Forwarding}

Data subscriptions may be inverted at routers into forwarding states that direct traffic from a rendezvous point or (one or several) sources to the receivers. These mechanisms construct a distribution tree for individual receivers and are known from PIM-SM/SSM for IP or Scribe on the overlay. 

\subsection{Distribution of Forwarding States}

Multicast forwarding state implementation at the control plane may decouple from the data plane by distributing membership independently. BIDIR-PIM \cite{rfc-5015} and BIDIR-SAM \cite{wsw-plddp-11} construct distributions trees by establishing forwarding states in separate control operations.

\subsection{Hybrid Push/Pull Schemes}

Data may be pushed to network edges using any of the routing schemes sketched above, and pulled by receivers using common unicast or tunnels. Such approaches are of particular deployment-friendliness and resemble current Content Distribution Networks. Content access may be simplified by carrying content replication servers as instantiators within the group name.

\subsection{Unicast-based Server Reflection}

The most simple approach to group communication is based on a single reflector server that forwards multicast data per unicast requests. Even though of limited scalability, this is the most common solution in deployment, and facilitates a trivial routing, whenever the reflector is named as instantiation point.

\section{Conclusions and Outlook}\label{sec:c+o}

In this paper, we have discussed the role of multicast communication in future name-oriented networks and identified the deficiencies of a pure unicast-based model. We thoroughly explored the aspects of group-oriented naming and demonstrated its potentials in group forming, routing, rendezvous and security.

 Even though only of illustrative value, we should mention our implementation of the name-oriented concept of an abstract multicast. Within the \hamcast\footnote{See {http://hamcast.realmv6.org for a current release.}} middleware, we are able to access groups on a very high level of abstraction \cite{mcsw-sumha-11,cmsw-mtgcg-11}, while providing an automated mapping to available communication technologies.  

In future work, we will extend our middleware to include prototypes of name-oriented network implementations and explore the gains of functionality and performance in practice.

\bibliographystyle{IEEEtran}
\bibliography{/bib/own,/bib/rfcs,/bib/ids,/bib/theory,/bib/layer2,/bib/internet,/bib/transport,/bib/overlay,/bib/ngi,/bib/vcoip,/bib/security}

\end{document}